\documentclass{ws-p8-50x6-00} 

\begin{document} 
 
\title{Goldstone boson--nucleon dynamics: \\ Working group summary and outlook} 
 
\author{Ulf-G. Mei{\ss}ner,$^\star$ Greg Smith$^\ddag$ \\(C\lowercase{onvenors})} 
 
\vspace{0.5cm}

\address{$^\star$Forschungszentrum J\"ulich, IKP (Th), D-52425  
J\"ulich, Germany \\ $^\ddag$TJNAF, Newport News, 
VA 23606, USA \\ 
E-mail: u.meissner@fz-juelich.de, smithg@jlab.org} 

\vspace{0.5cm}

\author{S.-I.~Ando, R.~Beck, V.~Bernard, B.~Borasoy, P.~B\"uttiker, S. Choi,
        P.J.~Ellis, J.~Goity, H.-W.~Hammer, D.L.~Hornidge, S.P.~Kruglov,
        B.~Kubis, R.~Lewis, R.A.~Lindgren,$^{\lowercase{x}}$ R.~Meier, 
        M.~Moj\v zi\v s, G.C.~Oades, J.A.~Oller,
        M.G.~Olsson, C.~Papanicolas,$^\dag$ M.M.~Pavan, M.L.~Pitt, 
        M.J.~Ramsey-Musolf, M.E.~Sainio, S.~\v{S}irca, R.P.~Springer, J.~Stahov, 
        R.~Tacik, C.~Unkmeir,$^\dag$ M.~Vanderhaegen\\
        (P\lowercase{articipants}) }

\footnotetext{$^\dag$ written contribution not received}

\footnotetext{$^x$ JLab proposal, not yet available }

 
\maketitle

\vspace{-8.cm}

{\hspace{6.4cm} {\tiny FZJ-IKP(TH)-00-27, JLAB-PHY-00-22}}

\vspace{8cm}

\abstracts{We highlight some of the recent results in chiral dynamics for 
systems with one nucleon/baryon presented at Chiral Dynamics 2000. We outline 
the most urgent experimental and theoretical challenges to be tackled in the 
coming years.} 
 
\vspace{0.2cm}

\section{Introduction} 
 
This working group was concerned with processes involving exactly one 
nucleon (baryon), addressing topics like pion--nucleon scattering, the 
$\pi$N sigma--term, all kinds of nucleon form factors, real and virtual Compton 
scattering, electromagnetic pion production, strangeness and so on. 
There has been considerable progress on the experimental as well as on 
the theoretical side in this domain of chiral dynamics. This was 
reported in a number of plenary talks as well as the working group 
contributions compiled in these proceedings. It is the main task of 
this summary talk to point out the directions of research which should 
be pursued in the coming years. This certainly is a highly subjective 
undertaking, nevertheless, the presentations given at this 
conference have highlighted a variety of clear--cut questions and 
problems to be addressed. These will be touched upon here.

\section{Theory: Status and perspectives} 
\label{sec:thy} 
 
\noindent {\bf Formal aspects:} The tool to perform the 
calculations in chiral perturbation theory (CHPT) or extensions thereof 
is an effective Lagrangian of the Goldstone bosons coupled to matter 
fields. General decoupling theorems~\cite{GZ} tell us that to leading 
order only the ground--state baryons should be included, the effect 
of baryon as well as meson resonances appears indirectly through the 
low--energy constants of the effective Lagrangian. This effective 
Lagrangian has been worked out to complete one--loop accuracy (fourth order) for the 
two--flavor case~\cite{mojzis} including renormalization. The precise treatment of 
the baryon fields is still under debate. The baryon mass scale 
complicates the power counting. This can be dealt with in two ways. 
The first solution, the so--called heavy baryon approach (HBCHPT), 
is by now standard and many processes have been studied in that 
framework and many interesting results have been obtained.  
It has the disadvantage that due to the strict expansion in 
the inverse of the baryon mass, in some cases the analytical structure 
of a given amplitude can be deformed. That happens e.g. in case of the 
nucleon scalar form factor or the isovector electromagnetic ones (as detailed 
below). In the first 
case, this has numerical consequences, in the second it does 
not.\cite{BKMff} An alternative approach has been discussed in the 
plenary talks by Becher and Leutwyler. It is based on a different 
regularization of the relativistic loop integrals (see also ref.\cite{ET}) 
and automatically fulfills all analyticity requirements in the 
low--energy region (the cut structure of the so regulated integrals 
can become incorrect for momenta outside the range where CHPT is 
valid).  The method is 
called infrared regularization (IR). Another advantage is the 
automatic resummation of all recoil corrections through the full Dirac 
propagator. There is, however, some ambiguity in treating the 
polynomial pieces stemming from the numerators in loop integrals.  
This very elegant and promising method certainly has to be scrutinized by a thorough 
investigation of many processes. Such a program is underway at various 
institutions. Much work has been done to extend 
effective Lagrangians to higher energies. Only in the case of 
including the spin-3/2 decuplet (the delta and its cousins)  
a truly systematic effective field theory has 
been formulated~\cite{HHK} after the pioneering work in 
ref.\cite{JMd} (based on the observation that the octet-decuplet splitting 
can be treated as another small parameter compared to the scale of chiral 
symmetry breaking). The inclusion of vector mesons still needs to be 
addressed in more detail (see refs.\cite{VM} for the present status), 
the problem to overcome is the non--conservation of boson number, which 
makes it difficult to formulate a consistent power counting.  
Furthermore, resummation methods have been used 
to not only improve convergence but also to generate (pseudo) bound 
states. This inevitably leads to some model-dependence, which can 
however be minimized by employing appropriate dispersion relation  
techniques.\cite{oller} 
This is particularly important for the studies of chiral dynamics with 
strange quarks, as discussed below, or to extend $\pi$N scattering into the 
region of the first and second resonances. 
 
\begin{figure}[htb] 
\begin{center} 
\epsfysize=2.75in  
\epsfbox{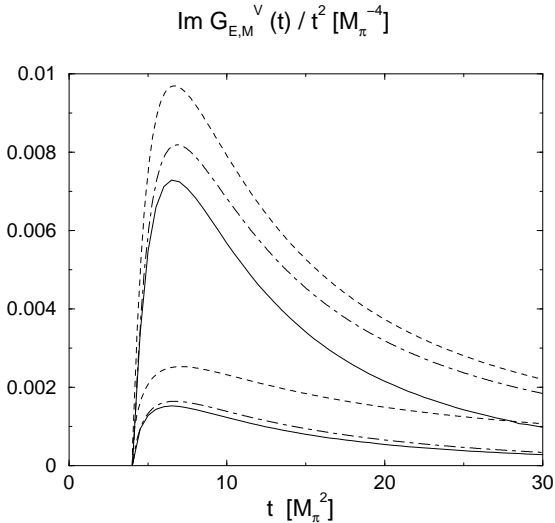} 
\end{center} 
\caption{Spectral distribution of the isovector electric and magnetic 
nucleon form factors weighted with $1/t^2$ calculated in heavy baryon 
CHPT. Shown are Im\,$G_M^V (t)/t^2$ (upper lines) and  
Im\,$G_E^V (t)/t^2$ (lower lines). The dot--dashed and dashed lines refer to the  
order $q^4$ and $q^3$ calculations, respectively. These results are very close 
to the ones based on a dispersion theoretical analysis with the $\rho$--meson 
contribution subtracted, as shown by the solid lines.} 
\label{fig:spectral} 
\end{figure}

\medskip 
\noindent {\bf Nucleon form factors:} The electromagnetic form 
factors are a good testing ground for certain aspects of the theory. 
First, the isovector radii are dominated by the anomalous threshold 
in the triangle diagram, which is exactly reproduced in the 
relativistic (IR) framework. Using HBCHT, one encounters a formal 
divergence. Its influence is, however, suppressed by phase space and 
one still obtains a decent description of the isovector spectral 
functions in HBCHT,\cite{BKMff} as shown in fig.\ref{fig:spectral}. 
 Second, from studies of the neutron charge form 
factor in HBCHT, one expects a bad convergence due to recoil effects. 
This expectation is borne out by the calculation using IR, it leads 
to a substantially  improved description of this much discussed observable.\cite{kubis} 
There is also some progress concerning the strange and bizarre (a.k.a. 
anapole) form factors, which are at the heart of some dedicated 
experimental programs at JLab, Bates and MAMI.  
A third order HBCHPT analysis of the strange proton form factors is 
available,\cite{HKM} but it again might prove fruitful to combine 
these with the dispersive results presented here.\cite{hammer} In addition, the 
anapole moment~\cite{musolf} and corresponding form factor pose a veritable 
challenge to theory, not so much from the chiral dynamics point 
(for calculations of the momentum dependence of the anapole 
form factor, see e.g.\cite{anaff}), 
but rather from the point of radiative corrections and 
parity--violating meson--nucleon interactions. A dedicated theoretical 
effort is needed to sharpen the tools to uniquely extract the physics 
hidden in the data of the existing and upcoming parity violation experiments. 
 
\medskip 
\noindent {\bf Pion--nucleon scattering:} Pion--nucleon 
scattering has long been recognized as a particular playground 
for chiral dynamics, triggered by the observation that the isoscalar S--wave 
scattering length vanishes in the chiral limit. 
In the last years, very  detailed and precise 
studies have been performed or are being performed in the heavy baryon 
as well as in the IR formalism. Much interest has focused on the so--called 
sigma term, which is nothing but the matrix element of the QCD 
quark mass term within proton states at zero momentum transfer. 
This quantity can only be obtained indirectly by various means 
- dispersion relations,\cite{sainio,stahov,pavan}  
sum rules\cite{olsson} (which let one express the sigma 
term via threshold parameters and calculable integrals) as well 
as by CHPT~\cite{paul} (or extensions therefore, see e.g.\cite{ellis}).
 In all cases, one needs experimental input as given 
in terms of various partial wave analyses (PWA). Unfortunately, only the 
by now outdated KA85 PWA passes all these tests. In addition, only 
recently a new calculation of electromagnetic corrections has become 
available~\cite{oades} and a full CHPT analysis including virtual photons is not 
yet completed. Since the precise determination of $\sigma (0)$ is only 
possible if one knows a variety of small corrections precisely, a more 
detailed look at isospin violation is certainly needed. First studies 
seem to indicate non--negligible effects, see refs.\cite{SMM}  
It is important to stress that the large range of values for the sigma 
term found in the literature appears to have narrowed down sizeably, as discussed 
in more detail in section~\ref{sec:exp}. Isospin 
violation is also of importance in the physical region close to 
threshold. Some phenomenological studies seem to indicate an 
astonishingly large effect in the S--wave (when comparing elastic 
scattering and CEX data via a triangle relation), but a complete 
CHPT calculation including virtual photons to settle this issue 
is not yet available but has to be done. So far, 
it has not been possible to obtain an accurate description 
simultaneously in the physical (threshold) region and inside the 
Mandelstam triangle. Clearly, a systematic marriage of the dispersive 
machinery with the chiral constraints has to be done. Such efforts are 
under way but are also very tedious, so it will take some time before 
a completely consistent and accurate picture of pion--nucleon 
scattering from the inside of the Mandelstam triangle to the low 
energy region of the available data has emerged. Furthermore, as discussed 
by Rusetsky, the effective field theory study of hadronic atoms offers 
the possibility of reliably extracting $\pi$N scattering lengths from 
pionic hydrogen and deuterium. Again, it is absolutely necessary to 
account for isospin violation. Finally, the analysis of the 
Goldberger-Treiman discrepancies in the octet favors a smaller $\pi$N 
coupling constant, $f^2 = 0.075$,\cite{goity} (which is consistent 
with most recent phase shift analysis of $\pi$N and NN scattering)  
if one assumes the standard scenario of chiral symmetry breaking (for an alternative 
view, see ref.\cite{stern}).

\medskip 
\noindent {\bf Electromagnetic meson production:} Threshold pion  
photo-- and electroproduction has been one of the 
cornerstones of testing chiral dynamics in the single nucleon sector. 
In particular, the production of neutral pions off protons and 
neutrons is very sensitive to explicit chiral symmetry breaking 
since the corresponding S--wave multipoles vanish in the chiral 
limit. This multipole also exhibits very clearly the unitary cusp due 
the opening of the secondary $\pi^+ n$ threshold. In addition,  
it was found many years ago that two particular 
combinations of the P--waves can also be predicted to high accuracy 
in a third order calculation. This needs to be sharpened by a complete 
fourth order calculation,\cite{BKMpw} in particular in view of the 
MAMI data reported here.\cite{beck} Also progress has been made for charged pion 
production and the inverse capture process.\cite{lewis} 
The case of neutral pion electroproduction is even more 
intricate, there appears to be a serious discrepancy between the 
HBCHPT calculation and the data as discussed by Merkel. Clearly, a 
more thorough look at the P--waves is also necessary here, in 
particular at their variation with the photon virtuality. The 
announced problem in the longitudinal S--wave multipole might well  
be a reflection of the insufficient treatment of the P--waves. 
Only after a 
complete understanding of these processes on the proton has been 
obtained, a truely qualitative analysis of extracting the 
corresponding neutron amplitudes from pion production off light nuclei 
will be feasible. Also, so far all studies have been performed within 
HBCHPT. While the analysis of photoproduction does not reveal any 
large recoil effects but rather is sensitive to some special and unique 
loop contributions, the IR formalism might allow to consider a 
larger range of photon virtualities in the electroproduction case. 
More work on these topics is urgently called for. 
 
\medskip 
\noindent {\bf Compton scattering:} Over the last few years, real (RCS) 
and virtual Compton scattering (VCS) of protons and neutrons has developed into 
another precision tool for not only testing chiral dynamics but also employing 
dispersive techniques.\cite{vand} Concerning the electromagnetic polarizabilities, it is 
now well established that major cancellations appear at NLO between the large 
delta and almost equally large $\pi$N loop corrections. This leads to a good 
agreement of theory with experiment at LO (which is a pure loop effect) and 
NLO for the proton. Unfortunately, the unsettled 
experimental situation concerning the neutron does so far not allow to draw 
any firm conclusion.\cite{horn} Much attention has also been paid to 
the spin sector, which is characterized by four 
spin--polarizabilities, but so far no direct experimental 
determinations of these fundamental quantities exist. This area will 
certainly gain importance in the coming years. 
Much attention has shifted to VCS, in particular since 
first data from MAMI have become available.\cite{roche} They agree amazingly well with the 
third order HBCHPT calculation at a fairly large photon virtuality.\cite{hemm} Clearly, 
a complete one--loop (fourth order) calculation is called for. A related topic 
is the momentum dependence of the DHG sum rule. The HBCHPT calculation is only 
applicable for small virtualities, but in case of the proton--neutron difference, 
where the resonance contributions drop out to a large extent, this range is somewhat 
larger.\cite{VB} Based on the experience obtained in the form factor calculation, using 
the IR formalism, this momentum dependence might be accurate 
up to a photon virtuality where pQCD is still valid. After many years of mumbling 
and talking,  a direct matching of the hadronic to the quark based description is  
in sight. Such a calculation is underway\cite{BHM} and its result is eagerly awaited.

\medskip 
\noindent {\bf Muon capture:} Ordinary (OMC) and radiative muon capture 
(RMC) on the proton is sensitive to the elusive pseudoscalar form factor  
of the nucleon and its associated coupling constant $g_P$. 
While the presently available data for OMC are consistent with the accurate CHPT 
prediction, the pioneering experiment on RMC performed at TRIUMF~\cite{TRI} lead 
to a value of $g_P$ exceeding the expection by 50\%. It was one of the 
highlights of this working group that two different groups~\cite{vero,ando}  
performed detailed analysis, 
which not only show that the method used by the TRIUMF group of simply rescaling 
the Born graphs $\sim g_P$ is inconsistent with what is known from OMC, but 
also that the combination of certain small effects related to the strong interactions  
as well as atomic physics can explain the measured photon spectrum using a coupling 
constant consistent with theoretical expectations. This is reminiscent of the 
sigma--term story that unfolded in $\pi$N scattering over the last decade. 
More precisely, the occupation numbers of the atomic structure in muonic atoms/molecules 
need to be carefully re-examined and a 
N$^2$LO calculation should be redone including all isospin breaking effects 
because of the sensitivity to the exact pion mass in the pion-pole contributions. 
 
\medskip 
\noindent {\bf Strange quarks:} Because of the fact that $m_s  
\sim \Lambda_{\rm QCD}$, it is not so obvious that the strange quark can be 
treated on the same footing as the light up and down quarks, for example 
one can entertain the possibility that the three-flavor quark condensate is 
much smaller than its two flavor cousin.\cite{orsay} Often, one finds rather 
sizeable kaon loop corrections which cast some doubt on the convergence of 
the chiral expansion. Some progress has been made e.g. in the discussion 
of the baryon masses or magnetic moments because the so--called reordering~\cite{mojzis} 
of the chiral series based on relating observables to a given order (i.e. 
performing a chiral expansion of the low energy constants) can improve the 
convergence dramatically.  This needs to be explored in more detail. 
In addition, the existence of (subthreshold) boundstates in certain channels of  
the SU(3) meson--baryon system necessitates the implementation of some resummation  
techniques. While quite a bit of progress has been made in the past years,\cite{mun,val}  
one should further minimize the model--dependence (which comes in e.g. via the 
regulator functions in the Lippmann--Schwinger equation). This can be done 
by using subtracted dispersion relations which has the further advantage that 
explicit resonance fields can be included by building up the crossed 
channel (left-hand) cuts in 
a perturbative manner.\cite{om2} In that way, one can address the question whether a  
particular resonance is ``pre-existing'' (corresponding to a quark model state) 
or is dynamically generated through the strong meson--baryon interactions. As 
an illustrative example how that can work, let us mention meson--meson 
scattering and the octet of scalar 
mesons, discussed e.g. in ref.\cite{oll} Other areas were three--flavor CHPT  
is very useful are the CP--violating sector of QCD,\cite{bora} or 
hypernuclear physics.\cite{springer}

\section{Experiment: Status and perspectives} 
\label{sec:exp} 
 
\medskip 
\noindent {\bf The $\pi$N $\Sigma$ Term -- Experiments:}  
With respect to the $\pi$N $\Sigma$ term, presentations were given 
describing new experiments particularly sensitive to the phases which 
determine $\Sigma_{\pi N}$, as well as new analyses of existing data 
which lead to improved determinations of $\Sigma_{\pi N}$. The new 
experiments which were described both capitalize on interference regions 
to heighten their sensitivity to the smaller partial waves.  
 
\medskip\noindent 
Meier described~\cite{meier} experiments at both TRIUMF and PSI to 
measure pion-proton analyzing powers at low energies. At TRIUMF the 
effort focussed on $\pi^-$p analyzing powers in the S-P interference 
region, which occurs at backward angles near 50 MeV. At PSI the focus 
has been on low energy $\pi^+$p analyzing powers near the 
Coulomb-nuclear interference region, although some $\pi^-$p data near 
the S-P interference region are also planned. Both experiments 
explored the region from roughly 50--100 MeV. By the time of the next 
chiral dynamics meeting, the data from both these experiments should 
be published. They should dramatically improve our understanding of 
the smaller low energy partial waves which have been difficult to 
accurately determine until now due to the normalization uncertainties 
which have plagued differential cross section measurements in the 
past at low energies. On top of that the analyzing power itself is an 
interference term, which gives it better sensitivity to small partial 
waves than the differential cross section. These small partial waves 
dominate at threshold and their accurate determination is crucial to our 
evaluation of $\Sigma_{\pi N}$.  
 
\medskip\noindent 
The other new experiment relevant to $\Sigma_{\pi N}$ was presented by 
Tacik.\cite{tacik}  He described measurements ongoing at TRIUMF which 
map out the angular dependence of the $\pi^\pm$p differential cross 
section at low energies right through the Coulomb-nuclear interference 
region. The experiment covers the kinematic regime from  
$\sim 6^\circ < \theta < 180^\circ$ and $15 < T_\pi < 
67$~MeV. Normalization uncertainties are mitigated by measuring  
$\mu^\pm$p scattering simultaneously.  
These data, which 
also should be available by the time of CD2003, provide a measure of 
Re(D$^+$) at t=0, where the $\pi$N amplitude D is the same amplitude 
which is directly proportional to $\Sigma_{\pi N}$ at the Cheng-Dashen 
point. In addition these measurements provide a direct measure of the 
$\pi$N scattering length a$^+_{0+}$, which along with it's P-wave 
counterpart a$^+_{1+}$ characterizes the usual determination of 
$\Sigma_{\pi N}$. Future plans of the TRIUMF CHAOS group are to 
explore the $\vec{H}(\pi^\pm , \pi^\pm \pi^+)n$ reaction near threshold. 
 
\medskip\noindent 
Taken together, and in light of the previously existing body of 
$\pi$N experimental data,  these new low energy $\pi$N  cross sections and analyzing 
powers will to a large extent complete the experimental information we 
require in order to determine $\Sigma_{\pi N}$. There are some 
relatively minor holes to be filled in and improvements to be made to be sure, but on the whole 
we will have about as complete an experimental picture of  low energy $\pi$N 
scattering as we can 
expect to ever have by the time of CD2003. It seems unlikely that any 
new experiments will come forward after that which could provide new information 
relevant to the sigma-term puzzle. On the other hand, it is equally 
clear that the two experiments discussed above will provide two crucial 
and at present missing pieces of this puzzle. It is at once 
unfortunate and exciting that these are probably the last pieces of 
the sigma-term puzzle that can be provided by $\pi$N scattering experiments.  
 
\medskip\noindent 
Having said that, it's important to point out that low energy $\pi$N  cross section and 
analyzing power measurements in the single charge exchange channel are 
still missing and are still important observables to 
measure. Like the extraordinarily precise pionic hydrogen atom  experiments,\cite{leise} 
they are primarily sensitive to isospin odd amplitudes, and  are thus 
to first order not especially relevant to $\Sigma_{\pi N}$, which is determined from 
isospin even amplitudes. 
Their importance lies primarily in the context of isospin violation, an 
extremely interesting topic in its own right. 
However, it is not yet clear how isospin violation may affect the 
value of $\Sigma_{\pi N}$ although first steps have been done in this direction.   
An effort to provide SCX observables at 
PSI seems to have stalled due to problems getting the appropriate 
$\pi^0$ detector to that laboratory. It is hoped that this 
bottleneck can be overcome by the time of the next meeting, and  
furthermore, that the impact of isospin violation on $\Sigma_{\pi N}$ will 
have been established theoretically by then as well. 
SCX at higher energies will be provided by PNPI, as discussed by 
Kruglov,\cite{pnpi} as well as spin rotation parameters which will be important 
in pinning down with better precision the higher partial waves which, 
we are beginning to suspect, play a greater role than previously 
thought in the determination of $\Sigma_{\pi N}$. 
 
\medskip 
\noindent {\bf The $\pi$N $\Sigma$ Term -- Analyses:}  
In addition to the new experiments reported at the meeting, there were 
several new analyses of $\pi$N scattering data presented which were 
used to deduce new values for $\Sigma_{\pi N}$. At the time of the 
last chiral dynamics meeting, only two analyses were available: 
KH80 and VPI/GWU. In a nutshell, each 
uses a different selection of the available body of experimental data 
to determine the $\pi$N partial wave amplitudes in the physical 
region. Each uses different techniques to extrapolate below threshold 
to the Cheng-Dashen point ($\nu=0, t=4M_\pi^2$) where the connection 
to $\Sigma_{\pi N}$ is made. It is widely agreed that the 
sub-threshold extrapolation machinery of the KH80 analysis is superior 
to that employed in the VPI/GWU analysis, although the sophistication 
of the VPI/GWU analysis is approaching  that of the original 
KH80 analysis  through the introduction of dispersion constraints. On the other hand, 
the data available to the KH80 analysis in 1980 was miniscule compared 
to what is presently available for use by the VPI/GWU analysis. In 
fact it appears that some of the most crucial experimental input 
used in the KH80 analysis was wrong, since it is at variance with all 
modern measurements performed at all three meson factories. 
In any case, the situation at CD97 was that the KH80 value for 
$\Sigma_{\pi N}$ was on the order of $64 \pm 8$ MeV, whereas the  VPI/GWU 
value was a whopping $92 \pm 3$ MeV, which corresponds to an $s 
\bar{s}$ content in the proton of over 25\%!  
 
\medskip\noindent 
So it was an important development in the field that at CD2000, 
several new determinations of $\Sigma_{\pi N}$ were presented, and in 
fact some degree of convergence was even observed.  
Stahov~\cite{stahov} reported the results of an analysis which 
combined fixed-t dispersion relations and interior dispersion 
relations. As input he chose the VPI/GWU partial waves (SP00) in the 
s-channel (the VPI/GWU analysis is far more succesful at reproducing 
experiment than KH80 is). Surprisingly, he found his results were 
relatively insensitive to the choice of VPI/GWU or KH80 partial wave 
input, a result at variance with the work of Sainio, discussed below. 
For the t-channel KH80 input was used. His 
result was $\Sigma_{\pi N}=72\pm 2$~MeV, where the error reflects only 
the uncertainty estimated for the extrapolation procedure. The error 
associated with the input partial waves is difficult to estimate and 
is one of the reasons we look forward to the next meeting when the 
results of the experiments discussed above will be available, 
presumably improving the partial wave input to analyses like these. 
Pavan~\cite{pavan} presented the latest VPI/GWU result, which changes 
in response to the increasing $\pi$N database as well as with 
improvements in their analysis, particularly with respect to their 
dispersion constraints. Their new result is $\Sigma_{\pi N}=84\pm 
5$~MeV, around 6~MeV lower than their previous result, but still  
significantly higher than KH80 or Stahov's result. 
Sainio~\cite{sainio} has reported at the previous meetings the result 
of an analysis based on six forward dispersion relations and partial 
wave input above a cutoff momentum of 185 MeV/c. In the past he has 
chosen KH80 phases as input, and got a result of $\Sigma_{\pi 
  N}=60$~MeV, consistent with the KH80 value of $\Sigma_{\pi N}=64\pm 8$~MeV. 
At CD2000 he reported a new result based on the same technique but 
using VPI/GWU phases as input (SP00). This moved his result to $\Sigma_{\pi 
  N}=93$~MeV, consistent with the previous VPI/GWU result of 
$\Sigma_{\pi N}=90$~MeV, and underscoring the sensitivity of these 
analyses to the partial wave input. It would seem obvious, given the 
vastly superior predictive power of the VPI/GWU phases relative to 
those of KH80, that a higher value of $\Sigma_{\pi N}$ is inescapable. 
Finally, a novel analysis was presented by Olsson~\cite{olsson} which 
used fixed-t dispersion relations to derive a new sum rule for 
$\Sigma_{\pi N}$ in terms of threshold paramters like the scattering 
lengths. His result of $\Sigma_{\pi N}=71 \pm 5$~MeV supports the 
trend to a higher value of $\Sigma_{\pi N}$ relative to the old KH80 
canonical value of 64 MeV.

\begin{figure}[htb] 
\begin{center} 
\epsfysize=4.8in  
\epsfig{file=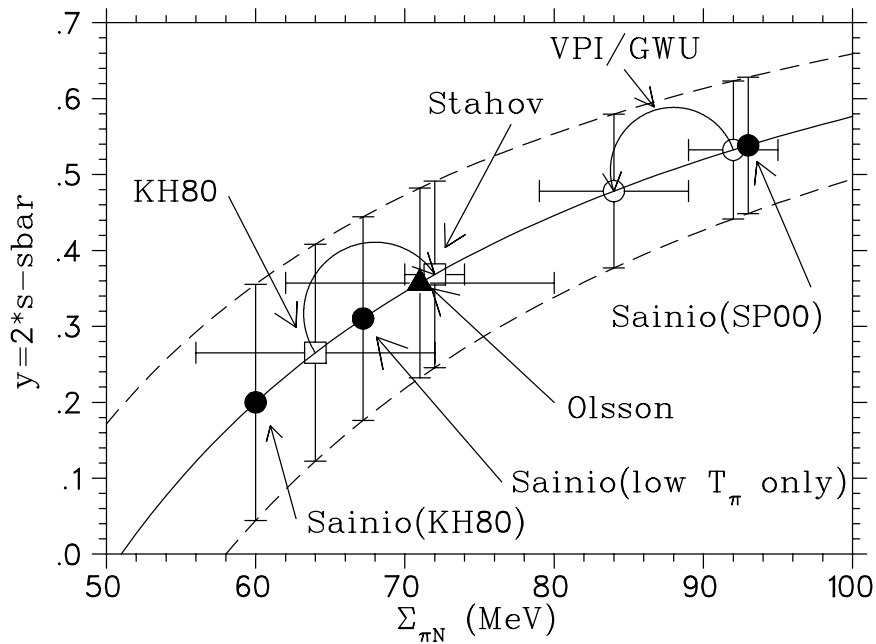, angle=90, width=4.5in} 
\end{center} 
\caption[gruel]{Values of the $\pi$N $\Sigma$ term reported at the workshop. 
The solid curve and it's dashed counterparts indicate the 
relationship between $\Sigma_{\pi N}$ and the $s\bar{s}$ content of 
the proton given by Borasoy and Mei{\ss}ner.\cite{ulf1} The points indicate the 
results of various analyses which lead to $\Sigma_{\pi N}$, 
discussed in the text.} 
\label{fig:sigma} 
\end{figure}

\medskip\noindent 
To summarize, the analyses of $\Sigma_{\pi N}$ seem to be converging 
on values between 71 and 84 MeV, implying that the $s\bar{s}$ content 
of the proton is 18--24\%, as depicted in fig.\ref{fig:sigma}. 
New experiments, especially those aimed at 
threshold parameters, should improve still further 
the convergence of the different 
$\Sigma_{\pi N}$ analyses by the time of the next chiral dynamics 
meeting.  
 
\medskip
\noindent {\bf Strange Form Factors:}  
Rather than probe the isoscalar amplitude connected 
to $\Sigma_{\pi N}$, a number of ongoing and planned experiments at electron 
facilities are probing the electric and magnetic form factors of the 
strange quark sea in the proton by measuring parity violation in ep 
scattering. Pitt~\cite{pitt} presented the status of the four 
experiments pursuing this line of research. So far only SAMPLE at 
Bates and HAPPEX at JLab have presented results for publication. By 
the time of the next chiral dynamics meeting, new results from both 
those collaborations as well as G0 at JLab and A4 at Mainz will be 
available. This will permit the separate extraction of $G^s_E$ and 
$G^s_M$ over a wide range of Q$^2$ (0.1 - 1.0 GeV/c$^2$) with 
overlapping results from 
different experiments at selected Q$^2$ values.  
The resulting insight ought to be one of the highlights of CD2003. 
Of related interest is the determination of the anapole moment of the 
nucleon, a topic discussed by Ramsey-Musolf.\cite{musolf} 
Experimental information on this fundamental parameter has 
been extracted from the SAMPLE experiment at Bates by combining the 
results of parity violating electron scattering asymmetries 
on hydrogen and deuterium targets. This, as well as new results in 
atomic parity violation, 
has stimulated theorists to revisit the anopole moment prediction 
using chiral perturbation theory. Unfortunately, the experimental 
measure of the anapole moment is at least an order of magnitude more 
challenging than the measure of the strange electromagnetic form 
factors. As a result, progress on this topic will be relatively slow 
in coming, and will no doubt be a topic of considerable interest at 
the next several chiral dynamics workshops.

\medskip 
\noindent {\bf Tests using electromagnetic probes:}  
Recent progress in our understanding of the axial form factor of the 
nucleon was summarized by \v{S}irca.\cite{sirca} For some time now there 
has been a puzzling discrepancy between results for the axial form 
factor determined on the one hand by quasi-elastic antineutrino 
nucleus scattering, and on the other by charged pion electroproduction 
on the proton. However, ancient CHPT work has actually predicted a 
difference ($\sim$5\%) in the axial mass determined each way. To test 
this prediction, the A1 group at Mainz has measured p(e,e$^\prime 
\pi^+$)n to high precision and has confirmed the predicted axial mass 
discrepancy. However, their results were acquired far enough from 
threshold that an effective Lagrangian model had to be employed rather 
than CHPT directly in the extraction of their result. As a 
consequence, measurements at lower Q$^2$ are planned, as well as 
measurements closer to threshold with a new device that should improve 
the reliability of their result in time for presentation at the next 
chiral dynamics conference. 
Measurements of the photon asymmetry in neutral pion photoproduction on 
the proton using TAPS at Mainz 
were described by Beck.\cite{beck} These data allow the separate 
determination of all (S- and) P-wave multipoles for the first time.    
Prior to this only the S-wave E$_{0+}$ and the P-wave multipoles P$_1$ 
and P$_{23}$ (a combination of the P$_2$ and P$_3$ multipoles) could 
be extracted from the available unpolarized cross sections. Only 
preliminary results were available at CD2000, but they confirmed the 
CHPT prediction for P$_2$. It will clearly be interesting to see 
the outcome of the stringent test of CHPT these data will 
constitute when 
the final results for all the multipoles are available at CD2003. By 
then we should also have the results of further experiments, now in 
the planning stage, such as that presented by Lindgren\cite{PET} to measure 
${\rm} H(\vec{e}, e^\prime p) \pi^0$ at Jefferson Lab. This 3 GeV 
coincidence experiment will cover the Q$^2$ region from 0.05 to 0.8 
GeV/c$^2$ and should permit the extraction of the S\&P wave multipoles 
very near threshold. The TAPS group is also planning further work, 
with higher intensity and with both beam and target polarized.  
 
\medskip\noindent 
Another fundamental property of the nucleon which can be predicted by 
CHPT is its electric and magnetic polarizability. Although these 
quantities are well measured for the proton, the neutron remains a 
formidable challenge. Hornidge~\cite{horn} described measurements 
of tagged photon elastic scattering from the deuteron, as well as 
quasi-free Compton scattering d($\gamma, \gamma^\prime$n)p at SAL 
which were performed to shed some light on the polarizibilities of the 
neutron. The elastic data have small statistical errors, but the 
extraction of polarizibilities is model dependent. Further theoretical 
guidance here is clearly required. In contrast, the quasi-free data 
minimized the model dependence but suffered from a lack of statistics. 
Given that further measurements at SAL are no longer possible,  it 
would be nice to see similar measurements performed somewhere else 
with more statistical significance by the time of CD2003. Happily, 
plans exist to measure the quasi-free channel at MAMI and the elastic 
channel at LUND. It is worth mentioning that a recent MAMI measurement
on the bound proton at backward angles
gives further credit to the idea of using quasi-free Compton
to determine the elusive neutron  polarizabilities.\cite{Wissmann} 
 
\medskip\noindent 
The spin structure functions g$_1$(x,Q$^2$) and g$_2$(x,Q$^2$) for the 
neutron as well as the Gerasimov-Drell-Hearn sum rule were 
investigated in an experiment described by Choi~\cite{choi} at 
Jefferson Lab. The ${\rm ^3\vec{He}(\vec{e},e^\prime )}$ reaction was 
studied from Q$^2$ of 0.03 to 1.1 GeV/c$^2$. The preliminary results 
reported at the workshop were in good agreement with older, much less 
precise data from SLAC and agreed well with a calculation by Drechsel, 
Kamalov, and Tiator. A follow-up experiment is in the works to pursue 
these measurements at much lower Q$^2$, where a reliable comparison to CHPT 
can be made, and where no information presently exists on the GDH integral.

\section*{Acknowledgments} 
The authors gratefully acknowledge the efforts and contributions of 
all the speakers in the Goldstone Boson--Nucleon working group. Their 
hard work and the high quality of their presentations resulted in a 
stimulating and productive working group. We also thank the organizers 
of the workshop, Aron Bernstein and Jos\'e Goity, as well as the hospitality 
and services provided by the hosting institution, Jefferson Lab.  
One of us (GRS) wishes to acknowledge the support of the US Department of Energy
under contract DE-AC05-84ER-40150.


\vspace{-0.3cm} 
 
\begin{thebibliography}{99} 
\bibitem{GZ}J. Gasser and A. Zepeda, Nucl. Phys. B174  (1980) 445. 
\bibitem{mojzis} M. Moj\v zi\v s, {\it Aspects of baryon CHPT}, these proceedings. 
\bibitem{BKMff}V. Bernard, N. Kaiser and Ulf-G. Mei{\ss}ner, 
  Nucl. Phys. A611 (1996) 429. 
\bibitem{ET}  P.J. Ellis and H.-B. Tang, Phys. Rev. C57 (1998) 3356. 
\bibitem{HHK}T.R. Hemmert, B.R. Holstein and J. Kambor, J. Phys. G24 
  (1998) 1831. 
\bibitem{JMd}E. Jenkins and A.V. Manohar, Phys. Lett. B259 (1991) 353. 
\bibitem{VM} 
G. Ecker et al., 
Nucl. Phys. B321 (1989) 311; G. Ecker et al., 
Phys. Lett. B223 (1989) 425;   
E. Jenkins, A. Manohar, and M. Wise, Phys. Rev. Lett. 75 (1995) 2272; 
B. Borasoy and Ulf-G. Mei{\ss}ner, Int. J. Mod. Phys. A11  
(1996) 5183;  
J. Bijnens, P. Gosdzinsky, and  P. Talavera, Nucl. Phys. B501 (1997) 495;   
JHEP 9801 (1998) 014;   Phys. Lett. B429 (1998) 111. 
\bibitem{oller} J.A. Oller 
    and Ulf-G. Mei{\ss}ner, {\it Chiral Unitary
    Approach to pion-nucleon scattering}, 
    these proceedings. 
\bibitem{kubis} B. Kubis, 
{\it Low-Energy Analysis of the Nucleon Electromagnetic Form Factors},
these proceedings. 
\bibitem{HKM}T.R. Hemmert, B. Kubis and Ulf-G. Mei{\ss}ner, Phys. Rev. C60 (1999) 
045501. 
\bibitem{hammer} H.-W. Hammer, 
{\it Strange Vector Form Factors of the Nucleon},
these proceedings. 
\bibitem{musolf} M.J. Ramsey-Musolf, 
{\it The Nucleon Anapole Moment},
these proceedings. 
\bibitem{anaff} C.M. Maekawa and  U. van Kolck, Phys. Lett. B478 (2000) 73.  
\bibitem{sainio} M.E. Sainio, 
{\it Pion-Nucleon Sigma-Term from SP00},
these proceedings. 
\bibitem{stahov} J. Stahov,
{\it Determination of the Sigma Term from the $\pi$N Scattering Data},
 these proceedings. 
\bibitem{pavan} M.M. Pavan, 
R.A. Arndt, R.L. Workman and I.I. Strakovsky,
{\it The Nucleon Sigma Term from the $\pi$N Partial Wave and
  Dispersion Relation Analysis},
these proceedings. 
\bibitem{olsson} M.G. Olsson, 
{\it A Threshold Sigma Term Method},
these proceedings. 
\bibitem{paul} P. B\"uttiker  
and Ulf-G. Mei{\ss}ner, {\it Dispersion Relations and ChPT}, 
these proceedings. 
\bibitem{ellis} P.J. Ellis,
{\it $\pi$-N from an Extended EFT}, 
these proceedings. 
\bibitem{oades} G.C. Oades, 
A. Gashi, G. Rasche, E. Matsinos and W.S. Woolcock,
{\it Electromagentic Corrections for $\pi^\pm$p Scattering},
these proceedings. 
\bibitem{SMM}Ulf-G. Mei{\ss}ner and S. Steininger, Phys. Lett. B419 (1998) 403; 
G. M\"uller and Ulf-G. Mei{\ss}ner,  Nucl. Phys. B556 (1999) 265. 
\bibitem{goity} J. Goity, 
{\it The Goldberger-Treiman Discrepancy and the Value of $g_{\pi NN}$},
these proceedings. 
\bibitem{stern} N.H. Fuchs, H. Szadijan and J. Stern, Phys. Lett. B238 (1990) 380. 
\bibitem{BKMpw}V. Bernard, N. Kaiser and Ulf-G. Mei{\ss}ner, forthcoming. 
\bibitem{beck} R. Beck and A. Schmidt, 
{\it Measurement of the Photon Asymmetry for the Production of Neutral
  Pions on the Proton in the Threshold Region},
these proceedings. 
\bibitem{lewis} R. Lewis, 
H.W. Fearing, T.R. Hemmert and C. Unkmeir, {\it Radiative Pion Capture
  in ChPT},
these proceedings. 
\bibitem{vand} M. Vanderhaegen, 
D. Drechsel, M. Gorchtein, A. Metz and B. Pasquini,
{\it Dispersion formalism for real and virtual Compton scattering},
these proceedings. 
\bibitem{horn} D.L. Hornidge, 
{\it Neutron Polarizability Experiments},
these proceedings. 
\bibitem{roche}J. Roche et al., Phys. Rev. Lett. 85 (2000) 708. 
\bibitem{hemm} T.R. Hemmert, B.R. Holstein, G. Kn\"ochlein and S. Scherer, 
Phys. Rev. Lett. 79 (1997) 22. 
\bibitem{VB} V. Burkert, nucl-th/0004001. 
\bibitem{BHM}V. Bernard, T.R. Hemmert  and Ulf-G. Mei{\ss}ner, forthcoming. 
\bibitem{TRI}G. Jonkmans et al.,  Phys. Rev. Lett. 77 (1996) 4512; 
D.H. Wright et al., Phys. Rev. C57 (1998) 373. 
\bibitem{vero} V. Bernard, 
T.R. Hemmert and Ulf-G. Mei{\ss}ner, {\it Ordinary and radiative muon
  capture on the proton},
these proceedings. 
\bibitem{ando} S.-I. Ando, 
{\it Muon Capture in CHPT},
these proceedings. 
\bibitem{orsay} B. Moussallam,  Eur. Phys. J. C14 (2000) 111;  
S. Descotes, L. Girlanda and J. Stern, JHEP 0001 (2000) 041; 
S. Descotes and J. Stern, Phys. Lett. B488 (2000) 274. 
\bibitem{mun} N. Kaiser, P.B. Siegel and W. Weise, Nucl. Phys. A594 (1995) 325. 
\bibitem{val} E. Oset and A. Ramos,  Nucl. Phys. A635 (1998) 99. 
\bibitem{om2} J.A. Oller and Ulf-G. Mei{\ss}ner, hep-ph/0011146. 
\bibitem{oll} J.A. Oller, hep-ph/0007349. 
\bibitem{bora} B. Borasoy, 
{\it The electric dipole moment of the neutron in chiral perturbation theory},
these proceedings. 
\bibitem{springer} R.P. Springer, 
{\it Hyperons and Hypernuclei},
these proceedings. 
\bibitem{meier} R. Meier, 
{\it New $\pi$N analyzing power measurements at PSI and TRIUMF},
these proceedings. 
\bibitem{tacik} R. Tacik, 
{\it The CNI Experiment at TRIUMF},
these proceedings. 
\bibitem{leise} H.C. Schr\"oder et. al, Phys. Lett. B469 (1999) 25. 
\bibitem{pnpi} S.P. Kruglov, 
{\it New results from PNPI and ITEP},
these proceedings. 
\bibitem{ulf1} B. Borasoy and Ulf-G. Mei{\ss}ner, Ann. Phys. 254 (1997) 192. 
\bibitem{pitt} M.L. Pitt, 
{\it The Nucleon's Strange Form Factors - Experiments},
these proceedings. 
\bibitem{sirca} S. \v{S}irca, 
{\it Axial form factor determinations at MAMI},
these proceedings. 
\bibitem{Wissmann}F. Wissmann et al., Nucl. Phys. A660 (1999) 232. 
\bibitem{choi} S. Choi, 
{\it Study of Neutron Spin Structure Functions at Low $Q^2$ with
  Polarized $^3$He},
these proceedings. 
\bibitem{PET}R.A. Lindgren et al., PET (Pion Electroproduction at Threshold) proposal,
in preparation.
\end{thebibliography}
\end{document}